\documentclass{svproc}
\usepackage{url}

\usepackage[utf8x]{inputenc}
\usepackage[T1]{fontenc}
\usepackage{graphicx}
\usepackage{longtable}
\usepackage{wrapfig}
\usepackage{rotating}
\usepackage[normalem]{ulem}
\usepackage{amsmath}
\usepackage{amssymb}
\usepackage{capt-of}
\usepackage[hidelinks]{hyperref}
\usepackage{tikz}
\usetikzlibrary{shapes,arrows,arrows.meta}
\usepackage{subfigure}
\usepackage{listings}
\usepackage{multicol}
\usepackage{natbib}

\definecolor{ruby}{rgb}{0.61,0.15,0.26}
\definecolor{write}{rgb}{0.61,0.15,0.26}
\definecolor{read}{rgb}{0.15,0.61,0.26}
\definecolor{ZIBgreen}{cmyk}{0.78, 0.11, 1, 0}
\definecolor{ZIByellow}{cmyk}{0.18, 0.38, 1, 0.07}
\definecolor{ZIBorange}{cmyk}{0, 0.7, 0.95, 0}
\definecolor{ZIBred}{cmyk}{0.25, 1, 1, 0.26}
\definecolor{ZIB1}{HTML}{00869E}
\definecolor{ZIB2}{HTML}{00458C}
\definecolor{FU}{HTML}{99CC00}
\definecolor{JULIA}{HTML}{9359A5}
\definecolor{NHR}{HTML}{FC1FAB}
\definecolor{GCF}{HTML}{8FB6D7}

\usepackage{pifont}
\usepackage{minibox}

\lstdefinelanguage{Julia}%
{morekeywords={abstract,break,case,catch,const,continue,do,else,elseif,%
    end,export,false,for,function,immutable,import,importall,if,in,%
    macro,module,otherwise,quote,return,switch,true,try,type,typealias,%
    using,while},%
  sensitive=true,%
  alsoother={$},%
  morecomment=[l]\#,%
  morecomment=[n]{\#=}{=\#},%
  morestring=[s]{"}{"},%
  morestring=[m]{'}{'},%
}[keywords,comments,strings]%

\begin{document}
\mainmatter              
\title{Vahana.jl - A framework (not only) for large-scale agent-based models}
\titlerunning{Vahana.jl}  
%
\author{Steffen Fürst\inst{1}\inst{2}\inst{3} \and Tim Conrad\inst{2} \and Carlo Jaeger\inst{3}\inst{4} \and Sarah Wolf\inst{1}\inst{3} }
\authorrunning{Steffen Fürst et al.} 
%
%
\institute{Freie Universit\"{a}t Berlin
  \email{steffen.fuerst@fu-berlin.de}
  \and
  Zuse Institut Berlin
  \and
  Global Climate Forum
  \and
  Beijing Normal University}

\maketitle              

\begin{abstract}
  Agent-based models (ABMs) offer a powerful framework for
  understanding complex systems. However, their computational demands often
  become a significant barrier as the number of agents and complexity
  of the simulation increase. Traditional ABM
  platforms often struggle to fully exploit modern computing
  resources, hindering the development of large-scale
  simulations.  This paper presents Vahana.jl, a high
  performance computing open source framework that aims to address
  these limitations. Building on the formalism of synchronous
  graph dynamical systems, Vahana.jl is especially well suited for
  models with a focus on (social) networks. The framework seamlessly
  supports distribution across multiple compute nodes, enabling
  simulations that would otherwise be beyond the capabilities of a single machine.
  Implemented in Julia, Vahana.jl leverages the interactive Read-Eval-Print Loop (REPL) environment, facilitating rapid model development and experimentation.
  
\end{abstract}
\section{Introduction}

Agent-based models (ABMs) offer a powerful framework for understanding
complex systems by simulating the interactions of individual agents.
ABMs are particularly useful in systems where complex dynamics emerge
from the behavior of individual entities, making them a powerful tool
in fields ranging from economics and social sciences to ecology and
public health
\citep{grimm2011,tesfatsion+judd2006_handbook,crooks2017}.

Over the last decade, the dramatic increase in both the availability
of large data-sets and computing power has created new opportunities
for social simulations, enabling researchers to address more complex
questions. However, most traditional agent-based modeling systems
struggle to exploit the full potential of these resources due to
limited scalability and the lack of use of high-performance computing
in the agent-based modeling community \citep{polhill+al2023}.

Historically, improvements in computational capabilities were driven
by increases in CPU clock speeds. In contrast, recent advancements
have shifted focus towards multi-core processors and compute clusters,
emphasizing parallel processing over faster single-core speeds. While
numerous platforms are available to support the development of ABMs
\citep{aber+al2017,kravaro+bassiliades2015}, only a limited number
support the parallelization of a single simulation across multiple
cores, and even fewer support the distribution of the simulation
across multiple nodes of a computer cluster
\citep{rousset+al2016}. This limited support for parallelization
hampers our ability to create large-scale simulations.

With the exception of FLAME \citep{coakley+al2015} and FLAME GPU
\citep{richmond+al2023}, existing HPC-capable ABM platforms primarily
use a geographical environment to distribute agents across processor
cores \citep{rousset+al2016,fuerst+geiges2018,cosenza+al2011}. While
effective for certain models, this approach is limiting when agent
interactions are not inherently tied to spatial locations. In such
cases, many ABMs can be more naturally expressed as graphs, where
vertices represent agents and edges represent relationships or
messages between them.

In this work, we present Vahana.jl, a new open-source high-performance
framework for development of large-scale agent-based models of complex
(social) systems, implemented in the Julia programming language
\citep{bezanson+al2012}.\footnote{Code: \url{https://github.com/s-fuerst/Vahana.jl}\\
  Documentation: \url{https://s-fuerst.github.io/Vahana.jl/stable/}}
It is based on a discrete dynamical systems formulation known as
synchronous graph dynamical system (SyGDS) \citep{kuhlman+al2011,
  adiga+al2018}.

Expressing a model as a SyGDS has the inherent advantage of enabling
parallelization within individual simulations.  Vahana.jl seamlessly
distributes simulations across multiple nodes of a computer cluster,
allowing for large-scale models that would be not feasible on a single
node. At the same time, it simplifies the process of model development
through integration with the Julia Read-Eval-Print-Loop (REPL),
including functions that allow to inspect the simulation's current
state directly within the REPL, facilitating rapid experimentation.

\section{Graph Dynamical Systems in Vahana.jl}
A graph dynamical system models how a system, represented as a finite
graph, evolves over time. Each vertex in the graph has a state, and
rules (vertex functions) determine how a vertex's state changes based
on the states of the vertex itself and the state of neighboring
vertices. In a Synchronous GDS (SyGDS), all vertex states are updated
simultaneously. A SyGDS can be interpreted as a generalized cellular
automaton \citep{gutowitz1991}, where the neighborhood is determined
by the network instead of the position of the cell in a grid.  In the
context of an ABM, the vertices represent the agents, so from this
point on, 'agents' and 'vertices' will be used interchangeably.

In our extended graph dynamic system (ExGDS), the agents can change
the graph structure themselves by adding or removing vertices and
edges and/or changing the state of the vertices and edges.  The
information available to an agent is limited to its own state, the
state of the edges where the agent is at the target position, and the
state of the agents at the source position of these edges.

Furthermore, Vahana.jl supports modeling spatial interactions through
discrete rasters, along with the concept of 'global' elements
connected to the entire system.

\subsection{Extending synchronous graph dynamical systems}

Within our ExGDS, the graph structure evolves alongside the states of
its elements. This means that in a simulation, each time step has a
unique \textbf{directed} graph, denoted as \(G_t=(V_t,E_t\)), where
the sets of vertices \(V_t\) and edges \(E_t\) may differ from the
previous steps.

In addition, we introduce the concept of vertex types, where
\(\Delta_A\) represents the set of all possible vertex types. A
mapping function, \(\pi_a: V_t \rightarrow \Delta_A\), maps each
vertex to a specific type \(\delta_a\). This enables the representation of
diverse entities within the simulated system, including individuals,
households, organizations, and more.

Likewise, we also introduce edge types, where \(\Delta_E\) represents the
set of all possible edge types. A mapping function,
\(\pi_e: E_t \rightarrow \Delta_E\), maps each edge to a specific type
\(\delta_e\). The different edge types represent diverse interactions
or relationships between agents. Self-loops and multiple edges (even
of the same edge type) are permitted between vertices.

For each agent- and edge type $\delta$ we have a different set of state
variables \(x_{\delta_,1}, \ldots, x_{\delta_,n}\), where each
\(x_{\delta_,i}\) is element of a (potentially) different set of
possible states \(X_{\delta_,i}\). So the state space of $\delta$ is
\(\Theta_{\delta} = X_{\delta_,1} \times \ldots \times
X_{\delta_,n}\), where \(\Theta_{\delta}\) can be also the empty set.

Each vertex type has a transition function
\(f_{\delta_a} : G'_{v,t} \rightarrow (V'_{v, t+1}, E'_{v, t+1})\).
In this transition function:
\begin{itemize}
\item \(G'_{v,t}\) is the 1-neighborhood of vertex \(v \in G_t\). This 
  includes $v$ itself, all its immediate neighbors at the source of the edges,
  and the edges where $v$ is on the target position.
\item The output is a graph \((V'_{v, t+1}, E'_{v, t+1})\) where $V'_{v, t+1}$ and $E'_{v, t+1}$ can be empty sets.
  The endpoints $v_s$ and $v_t$ of an $e \in E'_{v, t+1}$ must be vertices of $G'_{v,t} \cup V'_{v, t+1}$.
  Note that the case where $v \notin V'_{v, t+1}$ represents the event of an agent $v$ no longer exists in the simulation (e.g., due to death).
\end{itemize}

To construct \(G_{t+1}\), the graph at the next time step, we take
the following actions:
\begin{itemize}
\item For each vertex $v$ in the current graph $G_t$, we apply \(f_{\pi_a(v)}\). 
\item We then combine the results to form the
  complete graph at the next time step. This can be represented as:
  \(G_{t+1} = (\bigcup_{v \in V_t} V'_{v, t+1}, \bigcup_{v \in V_t} E'_{v, t+1})\).
\item Remove all edges with an endpoint that is not in $V_{t+1}$. 
\end{itemize}

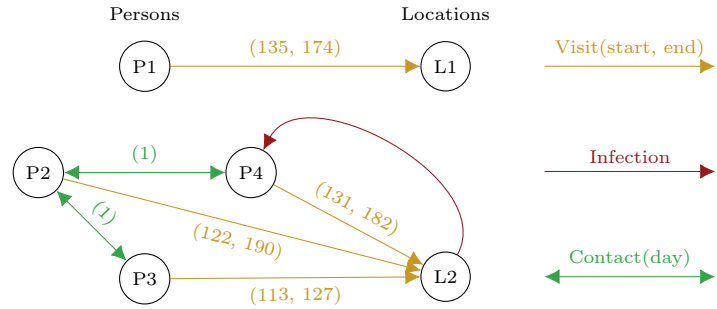
\begin{figure}[htbp]
  \centering
  \scalebox{1}{
\begin{tikzpicture}[-{Latex[length=2mm,width=2mm]}]
  \begin{scope}[font=\scriptsize]
  \node (Persons) {Persons};
  \node [right of=Persons, node distance=4cm](Locations) {Locations};
  \begin{scope}[shape=circle, node distance=2cm]
    \node[draw, below of=Persons, node distance=0.7cm] (P1) {P1};
    \node[draw, below left of=P1] (P2) {P2};
    \node[draw, below right of=P2] (P3) {P3};
    \node[draw, below right of=P1] (P4) {P4};
  \end{scope}
  \begin{scope}[shape=circle, node distance=2.8cm]
    \node[draw, below of=Locations, node distance=0.7cm] (L1) {L1};
    \node[draw, below of=L1] (L2) {L2};
  \end{scope}
  \begin{scope}[color=ZIBgreen, {Latex[length=2mm,width=2mm]}-{Latex[length=2mm,width=2mm]}]
    \path(P2) edge node[sloped,above]{\scriptsize{(1)}} (P4);
    \path(P2) edge node[sloped,above]{\scriptsize{(1)}} (P3);
  \end{scope}

  \begin{scope}[color=ZIByellow, -{Latex[length=2mm,width=2mm]}]
    \path(P1) edge node[sloped,above]{\scriptsize{(135, 174)}} (L1);
    \path(P2) edge node[sloped,below]{\scriptsize{(122, 190)}} (L2);
    \path(P3) edge node[sloped,below]{\scriptsize{(113, 127)}} (L2);
    \path(P4) edge node[sloped,above]{\scriptsize{(131, 182)}} (L2);
  \end{scope}

  \path(L2) edge[bend right=90, color=ZIBred] (P4);
  \begin{scope}[node distance=1.4cm]
  \node[right of=L2, node distance=1.2cm] (LegendContactL) {};
  \node[right of=LegendContactL, node distance=2.5cm] (LegendContactR) {};
  \path (LegendContactL) edge[color=ZIBgreen, {Latex[length=2mm,width=2mm]}-{Latex[length=2mm,width=2mm]}] node[above] {\scriptsize{Contact(day)}} (LegendContactR);
  \node[above of=LegendContactL, distance=4cm] (LegendInfectL){};
  \node[above of=LegendContactR] (LegendInfectR){};
  \path (LegendInfectL) edge[color=ZIBred, -{Latex[length=2mm,width=2mm]}] node[above] {\scriptsize{Infection}} (LegendInfectR);
  \node[above of=LegendInfectL] (LegendEndL){};
  \node[above of=LegendInfectR] (LegendEndR){};
  \path (LegendEndL) edge[color=ZIByellow, -{Latex[length=2mm,width=2mm]}] node[above] {\scriptsize{Visit(start, end)}} (LegendEndR);
  \end{scope}
  \end{scope}
\end{tikzpicture}
}
  \caption{An extended SyGDS representing a simplified version of an
    epidemiological model.
    Note that the locations are agents and that their transition functions determine which persons have been infected at this place and therefore generate the edges of type Infection. Since the locations have both persons in \(G'_{v,t}\), the source of the edge could also be the infector instead of the location. 
  }
  \label{fig:episim-network}
\end{figure}

We illustrate how a simple epidemiological model can be expressed as
such a graph in figure \ref{fig:episim-network}.  In this example, the
vertices stand either for individuals or for locations where
individuals interact and possibly spread the disease.

A key difference in Vahana.jl compared to other framework is how
agents interact with the system's structure. Rather than merely
updating directly the state of an agent, agents actively (re)construct
the relevant portions of a new graph at each time step. This approach
aligns with concepts like immutability from functional programming,
where the focus is on creating new values rather than modifying
existing ones. This simplifies parallelization and also excludes
artifacts that might arise from the order in which agents carry out
their transitions \citep{wolf+al2012d}. However, it requires that the
modeller explicitly define mechanisms to handle situations where
multiple agents compete for the same resource, instead of relying on
the "first-come-first-serve" behavior common in sequential systems
with random agent ordering.

It would be inefficient and cumbersome to require that even invariant
parts of the graph (such as stable relationship networks between
agents) be included in \((V'_{v, t+1}, E'_{v, t+1})\) for one vertex
$v \in V_t$. Therefore, in Vahana.jl, when a transition function is
applied, it is necessary to specify which types within the graph can
potentially change. Additionally, users can choose to retain all
existing agents or edges of a given type while only adding new ones
during the transition function.

\subsection{The spatial layer}

In Vahana.jl, (discrete) spatial information can be added via
n-dimensional rasters, where the raster cells are inserted as vertices
in the graph and a mapping from the Cartesian index of the underlying
space to the corresponding vertices can be accessed to determine the
vertex of a specific position. Since the cells are vertices of the
graph, they can be treated in the same way as other agents.

\begin{figure}[htbp]
  \centering
  \begin{tikzpicture}[scale=0.6, transform shape]

  \begin{scope}[color=black]
    \pgftransformcm{1}{0}{0.3}{0.35}{\pgfpoint{-2.5cm}{0cm}};
    \draw [gray!50,step=2.5cm] grid (12.5,10);

    Erstelle ein Raster von Kreisen
    \foreach \ii in {1,3,...,9} {
      \foreach \jj in {1,3,...,7} {
        \pgfmathsetmacro{\i}{1.25*\ii}
        \pgfmathsetmacro{\j}{1.25*\jj}
        \node[circle, draw, minimum size=1cm, color=FU] (node-\ii-\jj) at (\i, \j) {$l$};
      }
    }
    \foreach \i [evaluate=\i as \nexti using int(\i+2)] in {1,3,...,9} {
      \foreach \j [evaluate=\j as \nextj using int(\j+2)] in {1,3,...,7} {
        \ifnum \i < 9
          \draw[<->] (node-\i-\j) -- (node-\nexti-\j);
        \fi
        \ifnum \j < 7
          \draw[<->] (node-\i-\j) -- (node-\i-\nextj);
        \fi
      }
    }
  \end{scope}
  
  \begin{scope}[shape=circle, minimum size=1cm, color=ZIB2]

    \draw (2,6) node[draw] (S1) {};
    \draw (6,6) node[draw] (S2) {};
    \draw (10,6) node[draw] (S3) {};
    \draw (4,6) node[draw] (B1) {};
    \draw (8,6) node[draw] (B2) {};
  \end{scope}

  \path (13,6) edge[color=white] node[above,color=black] {\Large Edge Types:} (17,6);
  
  \begin{scope}
    \path (13,5) edge[<->] node[above] {\large Neighbor} (17,5);
  \end{scope}

  \begin{scope}[color=ZIB1, ->]
    \draw  (node-3-5) -- (S1);
    \draw  (node-5-5) -- (B1);
    \draw  (node-5-5) -- (S2);
    \draw  (node-7-1) -- (B2);
    \draw  (node-9-5) -- (S3);
    \path (13,4) edge[->] node[above] {\large Position} (17,4);
  \end{scope}

  \path (13,2) edge[color=white] node[above,color=black] {\Large Agent Types:} (17,2);
  \begin{scope}[shape=circle, minimum size=0.7cm]
    \draw (13.5,0.4) node[draw,color=ZIB1] {};
    \draw (14,0.4) node[color=ZIB1, anchor=west] {\large Agent};
    \draw (13.5,1.4) node[draw,color=FU] {};
    \draw (14,1.4) node[color=FU, anchor=west] {\large Location};
  \end{scope}
  
\end{tikzpicture}
  \vskip-5mm
  \caption{A graph with spatial information.}
  \label{fig:spatial}
\end{figure}
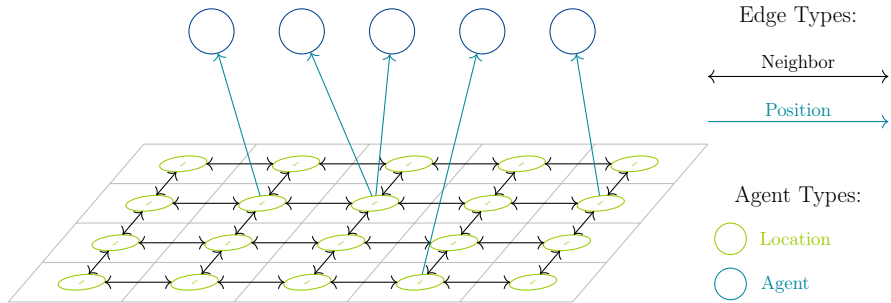

Vahana.jl provides specialized functions for creating and accessing
these spatial vertices and edges between them or between other
vertices/agents and the spatial vertices. These edges then represent
relationships such as "location \(l_1\) is a neighbor in the network
of location \(l_2\)" or "agent \(a\) is located at location \(l_1\)".
Figure~\ref{fig:spatial} shows an example of a graph with ``spatial''
vertices and edges, illustrating the kind of relationships Vahana.jl
can represent.

\subsection{The global layer}
\label{sec:orge8454b5}

The global layer contains vertices connected by edges to every other
vertex of the graph. This structure is useful for scenarios such as
providing model parameters to all or large sets of agents, collecting
information from agents to generate model outputs for visualization,
and aggregating information within the model to be relayed back to the
agents.

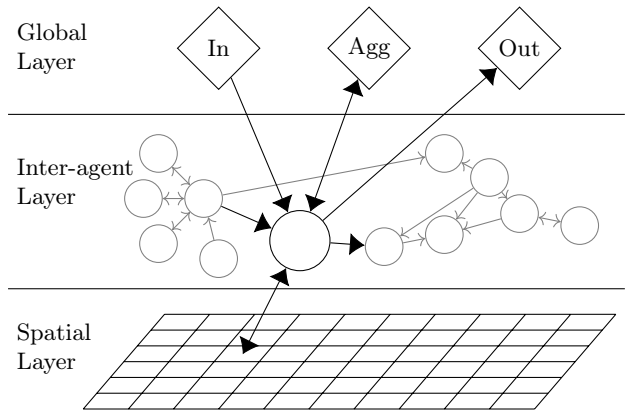
\begin{figure}[htbp]
  \centering
  \begin{tikzpicture}[scale=0.4]
  \draw (-5, 4) -- (15.5, 4);
  \draw (-5, 9.8) -- (15.5, 9.8);

  \begin{scope}[align=left,anchor=west]
    \draw (-5, 12) node { Global \\ Layer };
    \draw (-5, 7.5) node { Inter-agent \\ Layer  };
    \draw (-5, 2) node {  Spatial \\ Layer };
  \end{scope}
  
  \begin{scope}
    \pgftransformcm{1}{0}{0.3}{0.35}{\pgfpoint{-2.5cm}{0cm}};
    \draw [black,step=1.5cm] grid (15,9);
  \end{scope}

  \begin{scope}[shape=diamond, minimum size=1.4cm, font=\footnotesize]
    \draw (2,12) node  {In};
    \draw (7,12) node  {Agg};
    \draw (12,12) node {Out};
  \end{scope}

  \begin{scope}[shape=diamond, minimum size=1.1cm, font=\footnotesize]
    \draw (2,12) node[draw] (In) {};
    \draw (7,12) node[draw] (Agg) {};
    \draw (12,12) node[draw] (Out) {};
  \end{scope}

  \begin{scope}[shape=circle,  draw=black!50, minimum size=0.5cm, text=black!50, font=\small]
    \draw (2,5) node[draw] (A1) {};
    \draw (7.5,5.4) node[draw] (A3) {};
    \draw (9.5,5.8) node[draw] (A4) {};
    
    \draw (1.5,7) node[draw] (B1) {};
    \draw (11,7.7) node[draw] (B3) {};
    \draw (12,6.5) node[draw] (B4) {};

    \draw (0,8.5) node[draw] (C1) {};
    \draw (-0.5,7) node[draw] (C2) {};
    \draw (0,5.5) node[draw] (C3) {};
    \draw (9.5,8.5) node[draw] (C6) {};
    \draw (14,6.1) node[draw] (C7) {};

    \draw [<->] (C1) -- (B1);
    \draw [<->] (C2) -- (B1);
    \draw [<->] (C3) -- (B1);
    \draw [<-] (C6) -- (B3);
    \draw [<->] (C7) -- (B4);

    \draw [->] (A1) -- (B1);
    \draw [->] (A3) -- (A4);
    \draw [->] (B3) -- (A3);
    \draw [->] (B3) -- (A4);
    \draw [->] (B3) -- (B4);
    \draw [->] (B4) -- (A4);
    \draw [->] (B1) -- (C6);
  \end{scope}

  \begin{scope}[shape=circle, minimum size=0.8cm]
    \draw (4.7,5.6) node[draw] (A2) {};
    \draw [-{Latex[length=2mm,width=3mm]}] (B1) -- (A2);
    \draw [-{Latex[length=2mm,width=3mm]}] (A2) -- (A3);
    \draw [-{Latex[length=2mm,width=3mm]}] (In) -- (A2);
    \draw [{Latex[length=2mm,width=3mm]}-{Latex[length=2mm,width=3mm]}] (Agg) -- (A2);
    \draw [{Latex[length=2mm,width=3mm]}-] (Out) -- (A2);
    \draw [{Latex[length=2mm,width=3mm]}-{Latex[length=2mm,width=3mm]}] (A2) -- (2.8,1.82);
  \end{scope}
\end{tikzpicture}

  \caption{A graph example with all three layers of the ExGDS concept.
    Each rectangle in the spatial layer is a vertex of the graph.
    Edges between the layers are shown for a single vertex only.}
  \label{fig:layers}
\end{figure}

Figure~\ref{fig:layers} shows a graph with all three layers, where the
inter-agent layer is defined as the subgraph spanned by all vertices
that are neither a spatial nor a global vertex.

Please note that the global layer is a theoretical construct used to
demonstrate that the entire simulation state can be interpreted as a
single graph. In the Vahana.jl implementation, the global layer is
divided into \texttt{Parameters}, which remain constant throughout the
simulation and \texttt{Globals}. The Vahana.jl framework provides
functions to compute aggregations for distributed data and to
synchronously update the fields of the \texttt{Globals} instance.

\section{Implementation aspects}
This section explores the interplay of data distribution, performance
optimization, and reliability within Vahana.jl simulations.

\subsection{Data distribution}
\label{data-distribution}

In parallelized simulations, the distribution of data across multiple
processes is crucial for dealing with computationally intensive
problems.  Vahana.jl shields users from the complexities of low-level
distributed programming.  Under the hood, Vahana.jl leverages the
Message Passing Interface (MPI), a standardized and widely adopted
communication protocol for parallel and distributed computing.

In distributed Vahana.jl simulations, a fundamental design principle
is that each computational process knows only a subset of the graph
representing the current simulation state. Edges are always stored in
the process that manages the vertex at the edge's target, as this
agent must know of the edge's existence and can access its state
during transition functions.

Since an agent \(v\) can access its 1-neighborhood subset
(\(G'_{v,t}\)), the process that manages \(v\) must also have
knowledge of the states of the agents that are located at the source
of the incoming edges to $v$.  To optimize performance, Vahana.jl uses
MPI-3 shared memory to avoid redundant copies of agents residing on
the same compute node as the agent $v$. Only when the agents in
\(G'_{v,t}\) are managed by a different node, their state is
transferred to the process which is handling agent \(v\).

\subsection{Type Hints}
\label{sec:org305aa41}
In Vahana.jl, the specific code for accessing and managing the state
of the model is generated dynamically at runtime. This code generation
process can be influenced by so-called "hints" to improve performance.

Edge hints are optional traits that can be assigned to edge types.
These hints provide Vahana.jl with insights into how edges of a
specific type are utilized within transition functions.  Table
\ref{tab:edgehints} describes all edge hints and their effects. These
hints can be combined with one restriction: Due to implementation
details, the hints \texttt{SingleEdge} and \texttt{SingleType} can
only be used together if the hints \texttt{Stateless} and
\texttt{IgnoreFrom} are also given for the type.

\begin{table}[htbp]
  \caption{Edgetype Hints\label{tab:edgehints}}
  \centering
  \begin{tabular}{ll}
    Hint & Contract or Effect\\[0pt]
    \hline
    IgnoreFrom & Vahana.jl omits storing the source vertex ID.\\[0pt]
    IgnoreSourceState & The status of the agents at the source of the edge is not accessed.\\[0pt]
    SingleEdge & Each target vertex has at most one edge of this type.\\[0pt]
    SingleType & All target vertices are guaranteed to be of the same type.\\[0pt]
    Stateless & Only the source vertex ID is stored.\\[0pt]
  \end{tabular}
\end{table}

Hints play a crucial role in Vahana's code generation and runtime
behavior. They determine the specific implementation of data
structures such as the adjacency matrix and influence the generated
code for data access, promoting efficiency.  To illustrate the
influence of type hints on the generated code,
Fig.~\ref{fig:addedgecode} shows two different versions of the
\texttt{add\_edge!} function. Although simplified for clarity, they
represent the essence of the code after inlining and compiler
transformations and show how variations in hints can lead to
drastically different code.

\begin{figure}[htbp]
  \centering
  \begin{minipage}{1\linewidth}
    \begin{lstlisting}
# The "Knows" edgetype has no hints:
add_edge!(sim::Simulation, to::UInt64, edge::Edge{Knows}) =
  push!(get!(Vector{Edge{Knows}}, sim.Knows.write, to), edge)

# The "Knows" edgetype with all hints:
add_edge!(sim::Simulation, to::UInt64, edge::Edge{Knows}) = 
  sim.Knows.write[to & (2 ^ 36 - 1)] = true
    \end{lstlisting}
  \end{minipage}
  \caption{(Implementation of \texttt{add\_edge!} for two different combinations of hints.}
  \label{fig:addedgecode}
\end{figure}

The first implementation handles edge types without any optimization
hints, relying on operations such as dictionary lookups or array
appends. On an AMD Ryzen~9 7900X CPU, a call to this version requires
an average of 18.6\,ns (nanoseconds). In contrast, the second
implementation, optimized with all hints activated, reduces the
average execution time by more than 75\%, to just 4.5\,ns.

To utilize the MPI3 shared memory, which allows direct access to the
state of other agents within a single node to improve performance,
Vahana.jl stores agents in a contiguous block of memory. This requires
tracking the ``live'' state of the mortal agents to ensure memory
reuse within this contiguous block. When the \texttt{Immortal} hint is
applied to an agent type, it means that this agent type will persist
throughout the simulation, and therefore it is not necessary to track
the ``alive'' state.

These hints can be seen as contracts, that define the expected
behavior and interactions between agents, edges, and other components
of a model. For instance, in the context of edges, the
\texttt{SingleType} contract dictates that all agents connected by
such an edge at the target position must belong to the same
type. Violating this contract can lead to unpredictable and erroneous
simulation results.  To safeguard against such contract violations,
Vahana.jl by default performs checks that identify and flag potential
contract breaches. This allows for the correction of errors before
they cause simulation failures. However, once a model has been
thoroughly tested and verified, these checks can be disabled for
performance optimization.

\section{Case Studies and Performance Analysis}
\label{sec:org3b2df2a}
\label{sec:hkmodel}

In this section, we will examine two case studies that showcase how
Vahana.jl facilitates the construction and simulation of real-world
models.

All the results presented in this section were obtained from
simulation runs performed on the \textit{Lise} compute cluster at Zuse
Institute Berlin. Each node in this cluster is equipped with two Intel
Cascade Lake Platinum 9242 processors, each having 48 cores, and 384
GB of memory.

\subsection{Case Study 1: The Hegselmann-Krause (HK) Opinion Model}
\label{sec:orgdf9e609}
To assess the performance characteristics of Vahana.jl, we conduct an
analysis using the well-established Hegselmann-Krause (HK) Opinion
Model \citep{hegselmann+krause2002}.  This model describes a scenario
with a finite number of agents (\(n\)).  Each agent \(i\) holds an
opinion denoted by \(x_i(t)\), where \(t\) represents time that moves
forward in discrete steps. There is a ``confidence bound''
\(\epsilon>0\), so that agents only consider the opinions of others
that fall within a certain range of their own opinion.  At each time
step, all agents simultaneously update their opinions according to the
following rule:
\[ x_i(t+1) = \frac{1}{| \mathcal{N}_i(t) |} \sum_{j \in \mathcal{N}_i(t)} x_j(t) \]
\[ \textrm{where }\quad \mathcal{N}_i(t) = \{ j \in \{1, \ldots, n\}: \| x_j(t) - x_i(t) \| \leq \epsilon \} \]

In the classical HK model, each person can potentially be influenced
by all other persons. If we express this as a graph where the edges
indicate that one person's opinion is "visible" to another person, we
have a complete graph with an additional self-loop. The self-loop is
necessary as \(i \in \mathcal{N}_i(t)\). The complete implementation
of the HK model with Vahana.jl can be found in
Appendix~\ref{app:hkcode}.

Our first analysis addresses the overhead for using the Vahana.jl
framework, specifically the additional computational resources
required due to the framework's generality and abstraction layers by
comparing different implementations of the HK model. These include an
implementation utilizing Agents.jl \citep{datseris+al2022} , a direct
Julia implementation without additional packages, and one leveraging
Graphs.jl \citep{Fairbanks+al2021}.  For Vahana.jl, we compare two
implementations, one without type hints, and a second one that uses
the type hints \texttt{Immortal}, \texttt{Stateless} and
\texttt{SingleType}.

\begin{figure}[htbp]
  \centering
  \includegraphics[scale=0.13]{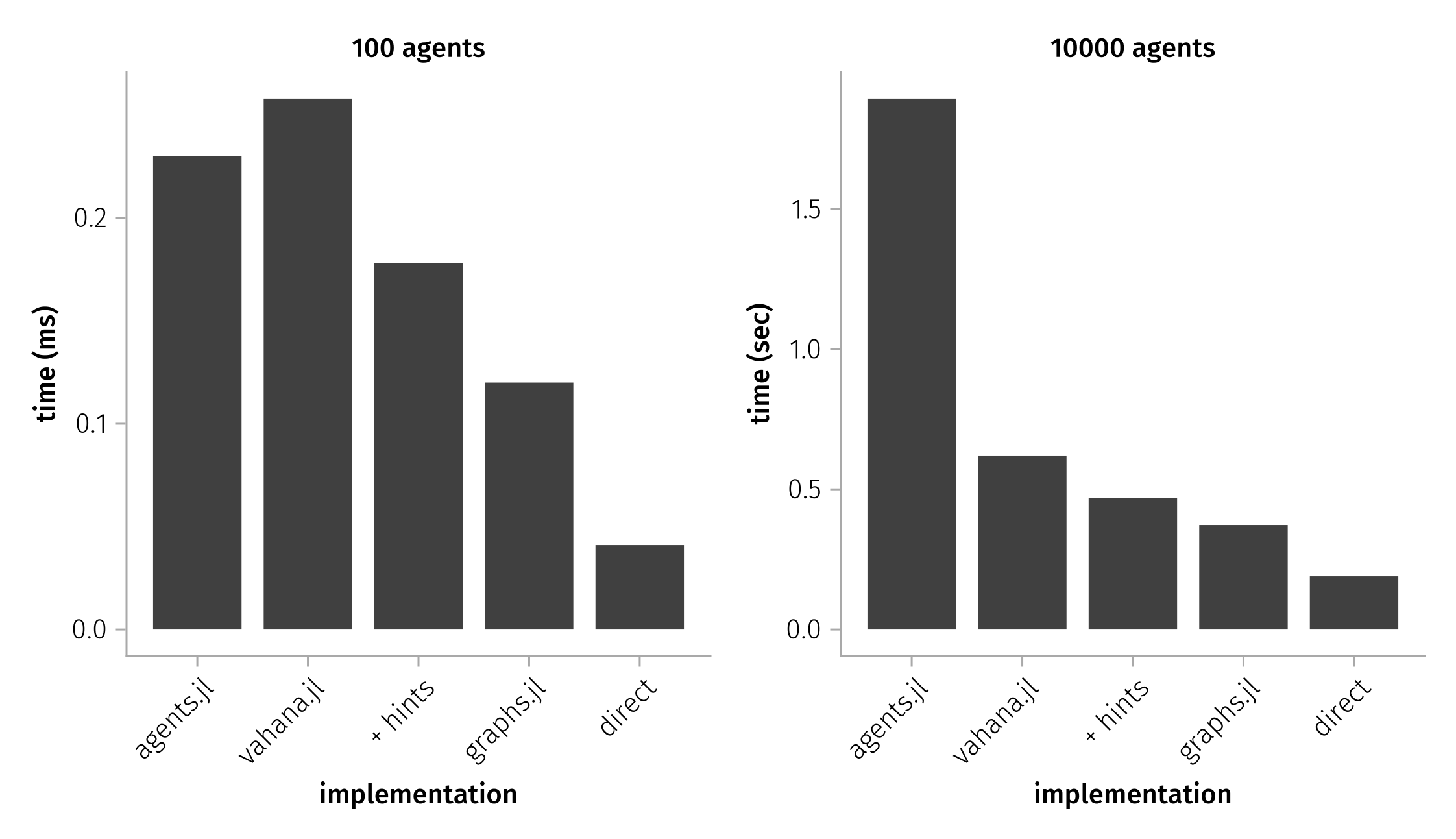}
  \caption{Comparison of serial performance of various implementations of the HK model. Note: The \texttt{+ hints} implementation is the Vahana.jl implementation with the \texttt{:Immortal, :Stateless} and \texttt{:SingleType} hints applied.}.
  \label{fig:other}
\end{figure}

Figure \ref{fig:other} shows how long a single iteration takes for a
simulation with 100 and 10,000 agents.  As expected, the performance
of a direct implementation cannot be achieved with Vahana.jl, but with
a larger number of agents, the performance overhead caused by the
framework is significantly reduced.  Notably, the direct
implementation and, to a lesser extent, the Agents.jl implementation
have an advantage as they bypass a potential graph structure
entirely. On the other hand, this means that in these implementations
the complete graph cannot be easily replaced by other graph
structures. To provide a more realistic comparison and also allow
exploring different graph structures without Vahana.jl, we
additionally included Graphs.jl. This reduces the difference
significantly. With 10,000 agents, the Vahana.jl implementation with
hints is only 25\% slower than the Graphs.jl implementation, and
significantly outperforms the Agents.jl implementation.

For the following performance analysis, we consider two additional
graph structures: regular graphs, where each vertex has the same
number of edges, and clique graphs, which consist of numerous
equal-sized cliques where only a single vertex from each clique is
connected to the neighboring cliques.  As explained in
Sect.~\ref{data-distribution}, the graph must be partitioned if a
simulation is to be processed in parallel, and simulations that use
multiple nodes of a computer cluster may require transferring the
state of agents between nodes. The amount of communication depends on
the structure of the graph, as it depends on the weight of cut edges,
where the weight depends on the size of the edge state.

\begin{figure}[htbp]
  \centering
  \includegraphics[scale=0.13]{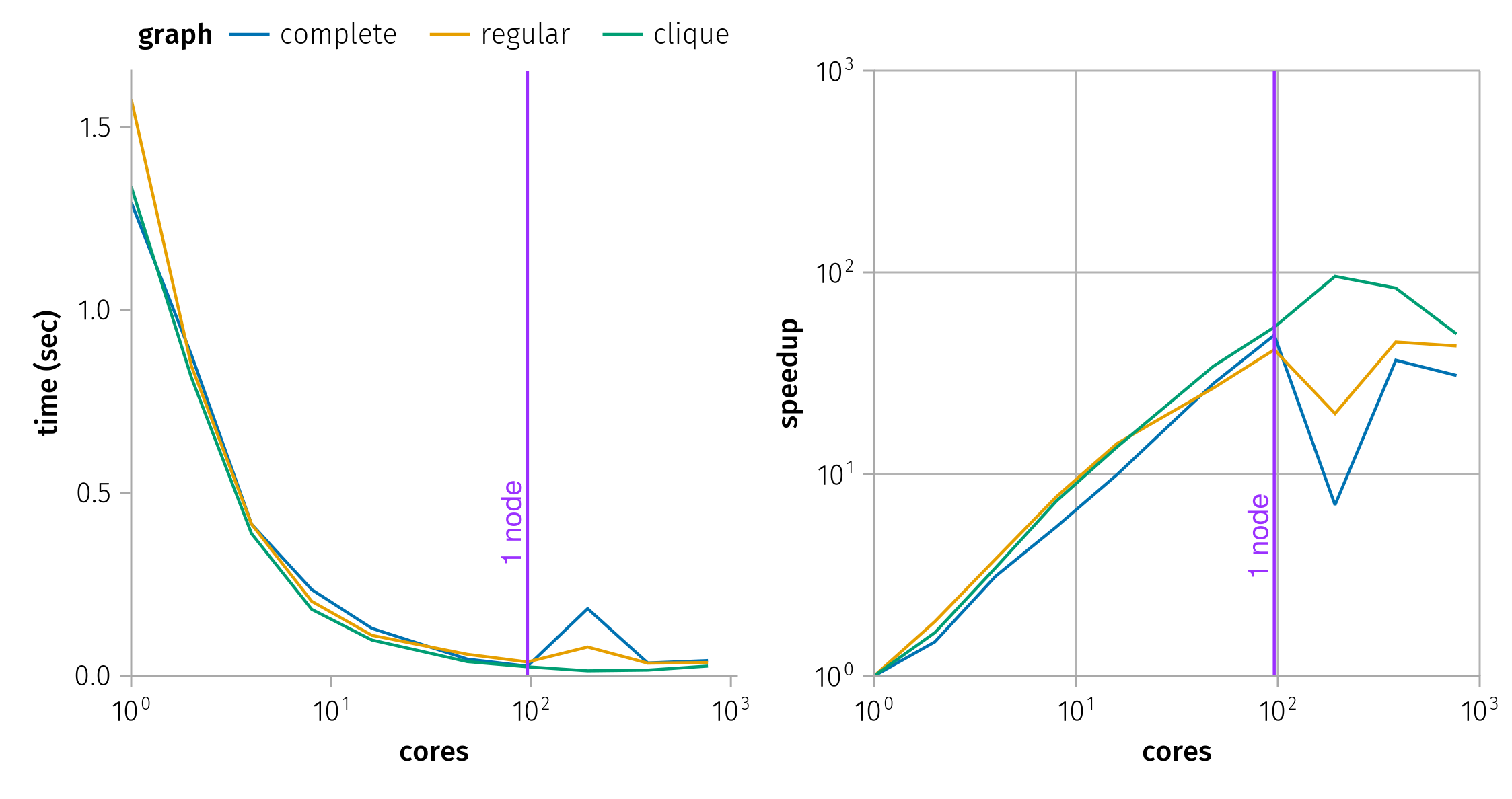}
  \caption{Scalability of the HK model with different graph structures.}
  \label{fig:scalability}
\end{figure}

The plot on the left of Fig.~\ref{fig:scalability} presents the
overall simulation runtime, while the figure on the right illustrates
the speedup achieved during transition steps disregarding
startup/compilation time. Ideally, the speedup would scale linearly
with the number of cores. This is somewhat true for the regular and
clique structures up to 16 cores. Even scaling up to a full node (96
cores), with values between 41 and 54, the speedup is quite good.

However, as expected, a clear difference in the scalability of the
different graph structures becomes apparent when we extend the
simulation to multiple nodes.  This is mainly due to the increasing
cost of accessing the opinion state of agents on a different node. For
a complete graph, if we distribute the agents evenly across the
computational processes, on average a fraction of
\(\frac{\#nodes - 1}{\#nodes}\) are edges where the agents at the
source and destination position of an edge are on different nodes, no
matter how the agents are partitioned. In the regular case, a good cut
can greatly reduce this number, and in a clique graph there is only
one edge per node that crosses the node boundaries if the number of
cliques can be divided by the number of (cluster) nodes.

\subsection{Case Study 2: The MATSim-Episim Model}
MATSim-Episim is a large-scale agent-based epidemiological model that
integrates a person-centric human mobility model with a mechanistic
model of infection and disease progression
\citep{10.1371/journal.pone.0259037}. It builds on the multi-agent
transport simulation (MATSim) \citep{horni+al2016}, an activity-based,
extensible, multi-agent simulation framework implemented in
Java. During the COVID-19 pandemic, this model was regularly used to
advise the German federal government (e.g. \citet{11303_12977}) by
providing reports containing detailed evaluations of the impact of
various interventions, such as reductions in out-of-house activities,
mask usage, and vaccinations.  Each report required thousands of
simulations based on mobile-phone data for Berlin and Cologne,
Germany, considering the activities of 25\% of a city's population.

We reimplemented the MATSim-Episim model described in
\citet{mueller+al2020} with
Vahana.jl\footnote{\url{https://git.zib.de/sfuerst/vahana-episim/}}
and tested the scalability for two different scenarios. In addition to
the Berlin scenario used for the reports with 1,013,973 agents,
889,958 locations and 6,814,215 daily location visits, we also
extended the area to the whole of Germany, again with 25\% of the
population, or more precisely with 16,334,073 agents, 11,538,167
locations and 91,534,225 daily location visits.

\begin{figure}[htbp]
  \centering
  \includegraphics[scale=0.13]{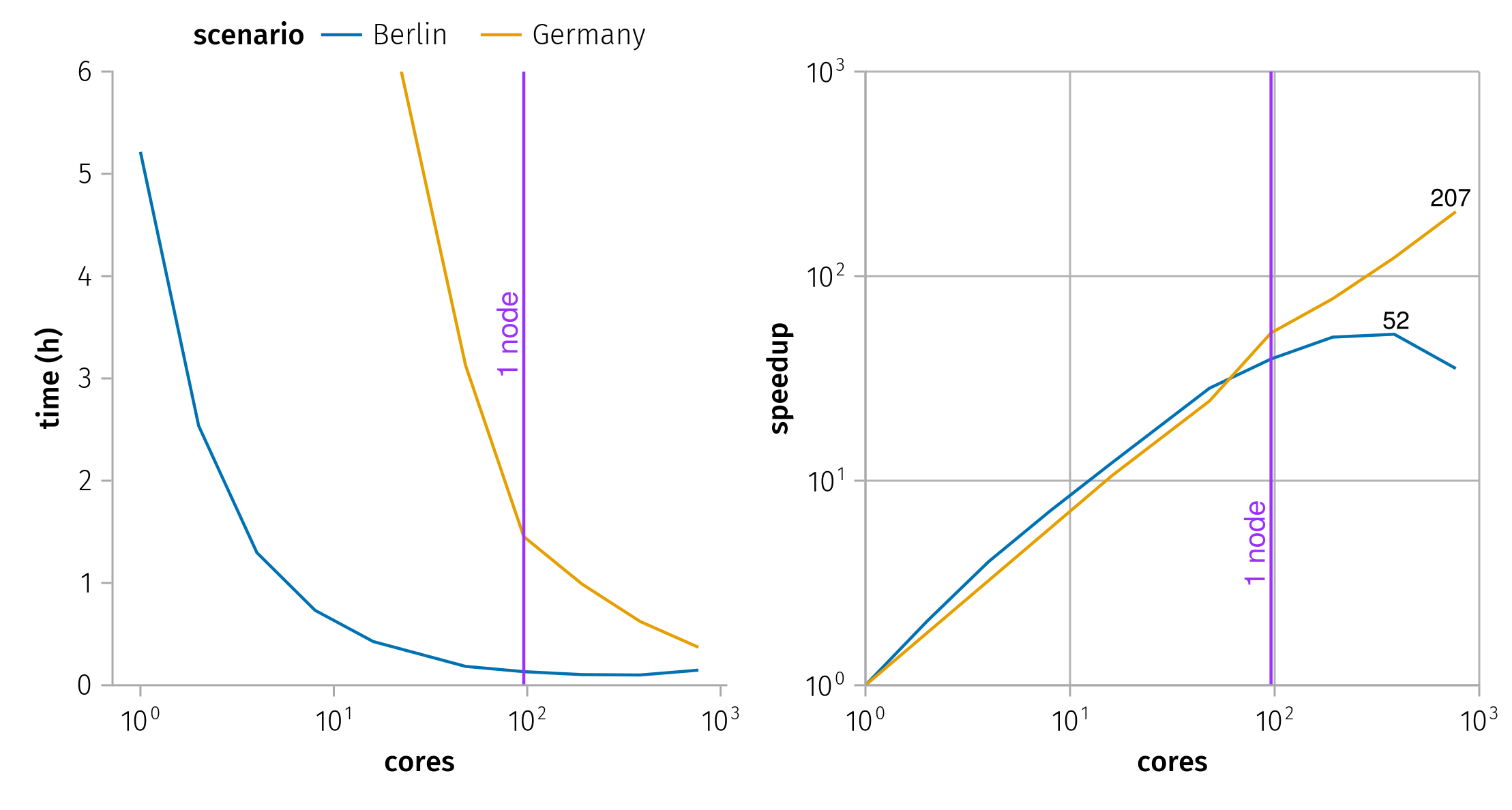}
  \caption{Scalability of the MATSim-Episim model.}
  \label{fig:matsim_scalability}
\end{figure}

The results are shown in Fig.~\ref{fig:matsim_scalability}. The plot
on the left presents the overall runtime for simulating 1 year, while
the figure on the right illustrates the speedup achieved.  We see that
the problem size for the Berlin scenario is not large enough to make a
distribution to multiple nodes on the \textit{Lise} compute cluster
reasonable.  This is different for the Germany scenario, where we
achieve an impressive reduction to 22.5\,minutes with 8 nodes, which
is 207 times faster than the 3.2\,days needed to calculate the Germany
scenario with a single thread.

\begin{figure}[htbp]
  \centering
  \includegraphics[scale=0.13]{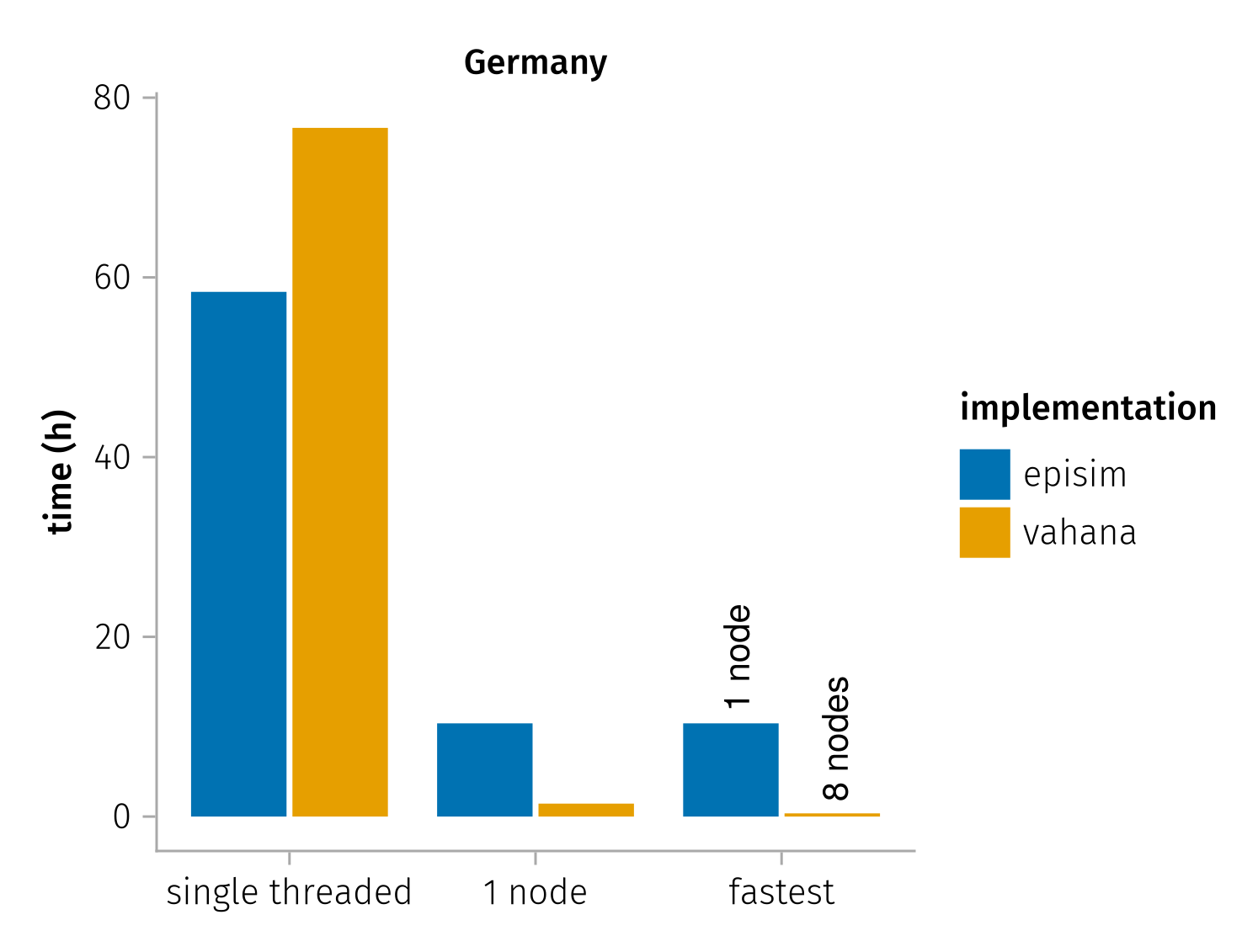}
  \caption{Performance comparition with original Java implementation.}
  \label{fig:episim_compare}
\end{figure}

Figure~\ref{fig:episim_compare} presents a performance comparison
between our Vahana.jl implementation and the original Java-based
MATSim-Episim for the Germany scenario. This should be viewed with
some caution, as the version from the MATSim-Episim
repository\footnote{\url{https://github.com/matsim-org/matsim-episim-libs}}
mentioned in the paper was used to measure the runtime, but it was not
checked whether additional features not described in the paper were
implemented there. Note that MATSim-Episim utilizes Java Parallel
Streams for its most computationally intensive loops, restricting
parallelization to a single node (\cite{fuerst+rakow2022}).  Despite
the slower single-thread performance of Vahana.jl compared to the Java
version, it shows remarkable advantages when parallelized. When using
a single node, Vahana.jl outperforms the original MATSim-Episim
implementation by a factor of 7. When scaled to 8 nodes, this
advantage increases to a factor of 28.

\section{Conclusion}
In this paper, we introduced Vahana.jl, an ABM framework that supports
simulations designed for parallel execution at various scales, from
single computers to large-scale HPC environments, while simplifying
the development of simulations designed for parallel execution.

We have outlined the theoretical foundations and design principles of
Vahana.jl, highlighting its unique approach compared to existing ABM
frameworks. This approach has been proven capable of facilitating the
development of complex ABMs, as demonstrated through the
epidemiological use-case, based on the reimplementation of the
MATSim-Episim model.

Our performance analyses demonstrate that Vahana.jl is a compelling
choice for researchers, particularly when network interactions between
agents are crucial. Notably, it enables researchers to tackle
significantly larger and more complex problems, thereby enhancing the
scope and potential of their simulations.

Vahana.jl is not limited to large-scale simulations; it is also a
powerful tool for researchers studying dynamics of networks,
regardless of scale. Its design philosophy prioritizes network
interactions and relationships, making it a valuable tool for various
research applications.

Future work will focus on further performance improvements,
particularly through optimizations tailored to common patterns and
tasks that arise specifically within the Vahana.jl framework. By
directly supporting these patterns, we aim to not only simplify model
implementation but also unlock additional performance gains for users.

\vskip1em
\footnotesize{\textbf{Acknowledgments} This research has been partially funded under Germany’s Excellence Strategy, MATH+: The Berlin Mathematics Research Center (EXC-2046/1), project no. 390685689} and by the European Union's Horizon Europe research and innovation program [grant agreement number 101059498—eco2adapt].

\bibliography{vahana}
\bibliographystyle{abbrvnat}

\newpage
\appendix
\lstset{%
  language         = Julia,
  keywordstyle     = \color{blue},
  basicstyle=\footnotesize,
  numbers=none,
  numberstyle=\tiny\itshape,
  frame=none,
}

\section{The HK model implemented with Vahana.jl}
\label{app:hkcode}
\scalebox{0.84}{
\begin{lstlisting}
using Vahana, Accessors
import Statistics: mean
import Graphs: SimpleGraphs

# define all types
struct Person opinion::Float64 end
# the "Knows" edges determine which agents can influence an agent
struct Knows end 
struct Params epsilon::Float64 end 

# create a simulation with a confidence bound of 0.2
const sim = ModelTypes() |>
    register_agenttype!(Person) |>
    register_edgetype!(Knows) |>
    create_model("Hegselmann-Krause") |>
    create_simulation(Params(0.2), nothing)

# we use the Graphs.jl package to construct a complete graph with
# 50 persons and add this graph to the Vahana graph.
const ids = add_graph!(sim,
                       SimpleGraphs.complete_graph(50),
                       # construct the person with an random opinion
                       _ -> Person(rand()),
                       _ -> Knows());

# the complete graph does not include a edge for the person to itself,
# so we add those in a second step
foreach(id -> add_edge!(sim, id, id, Knows()), ids) 

# `finish_init!` must be called before we can apply transition functions.
# E.g. in a parallel simulation, at this point the constructed graph will
# be partitioned and distributed to the different processes
finish_init!(sim)

# `transition` is called for each person, `self` is the state
# of the person, and `id` is needed to access other parts of the graph
function transition(self, id, sim)
    # get a vector with the opinions of the other Persons who are
    # connected to the Person `self` via edges of type `Knows`
    opinions = map(s -> s.opinion,
                   neighborstates(sim, id, Knows, Person))
    # remove all opinions that are outside of the confidence interval
    accepted = filter(opinions) do other_opinion
        abs(other_opinion - self.opinion) < param(sim, :epsilon)
    end

    # set the new opinion to the mean value of all accepted opinions
    @set self.opinion = mean(accepted)
end

# run 100 steps (= apply the transition function 100 times)
for _ in 1:100
    # the following can be read as: the transition function 'transition' is 
    # called for agents of type Person, has access to the state of Persons
    # and edges of type Knows and will create only Persons
    apply!(sim, transition, Person, [ Person, Knows ], Person)
end
\end{lstlisting}}

\end{document}